\documentclass[pdflatex,sn-mathphys-num]{sn-jnl}

\usepackage{graphicx}%
\usepackage{multirow}%
\usepackage{amsmath,amssymb,amsfonts}%
\usepackage{amsthm}%
\usepackage{mathrsfs}%
\usepackage[title]{appendix}%
\usepackage{xcolor}%
\usepackage{textcomp}%
\usepackage{manyfoot}%
\usepackage{booktabs}%
\usepackage{algorithm}%
\usepackage{algorithmicx}%
\usepackage{algpseudocode}%
\usepackage{listings}%

\usepackage{url}
\usepackage{paralist}
\usepackage{tikz}
\usepackage{float}
\usepackage{placeins}
\usepackage{makecell}
\usepackage{array,ragged2e,makecell}
\usepackage{tabularx}
\newcolumntype{Y}{>{\centering\arraybackslash}X}
\usepackage{fancyhdr}

\newcolumntype{P}[1]{>{\RaggedRight\arraybackslash}p{#1}}

\usetikzlibrary{arrows.meta, positioning, shapes.arrows, shapes.geometric}

\theoremstyle{thmstyleone}%
\theoremstyle{thmstyletwo}%

\theoremstyle{thmstylethree}%

\begin{document}

\title[Ethical AI Framework of the Future]{Building the Ethical AI Framework of the Future: From Philosophy to Practice}





\author*[1,2]{\fnm{Jasper Kyle} \sur{Catapang}}\email{catapang.jasper.kyle.y0@tufs.ac.jp}
\affil*[1]{\centering
\orgdiv{Graduate School of Global Studies}\par
\orgname{Tokyo University of Foreign Studies},
\orgaddress{\city{Tokyo}, \country{Japan}}
}
\affil*[2]{\centering
\orgdiv{AI Product Development Division}\par
\orgname{Money Forward, Inc.},
\orgaddress{\city{Tokyo}, \country{Japan}}
}

\abstract{
Artificial intelligence pipelines—spanning data collection, model training, deployment, and post-deployment monitoring—concentrate ethical risks that intensify with multimodal and agentic systems. Existing governance instruments, including the EU AI Act, the IEEE 7000 series, and the NIST AI Risk Management Framework, provide high-level guidance but often lack enforceable, end-to-end operational controls. This paper presents an ethics-by-design control architecture that embeds consequentialist, deontological, and virtue-ethical reasoning into stage-specific enforcement mechanisms across the AI lifecycle. The framework implements a triple-gate structure at each lifecycle stage: Metric gates (quantitative performance and safety thresholds), Governance gates (legal, rights, and procedural compliance), and Eco gates (carbon and water budgets and sustainability constraints). It specifies measurable trigger conditions, escalation paths, audit artefacts, and mappings to EU AI Act obligations and NIST RMF functions, enabling integration with existing MLOps and CI/CD pipelines. Illustrative examples from large language model pipelines demonstrate how gate-based controls can surface and constrain technical, social, and environmental risks prior to release and during runtime. The framework is accompanied by a preregistered evaluation protocol that defines ex ante success criteria and assessment procedures, enabling falsifiable evaluation of gate effectiveness. By translating normative commitments into enforceable and testable controls, the framework provides a practical basis for operational AI governance across organizational contexts, jurisdictions, and deployment scales.
}




\maketitle

\thispagestyle{fancy}

\fancyhf{}
\fancyhead[R]{%
\footnotesize
Accepted in \textit{AI and Ethics} (2026) \href{https://doi.org/10.1007/s43681-026-01003-8}{here}
}
\fancyfoot[C]{
1
}

\section{Introduction}
The rise of Artificial Intelligence (AI) has reshaped Natural Language Processing (NLP) and Information Retrieval (IR), powering applications from search and summarization to conversational agents and retrieval-augmented generation (RAG) \cite{Lewis2020}. Large Language Models (LLMs) currently exemplify this transformation, yet the field is evolving toward multimodal foundation models, agentic AI systems, and hybrid neuro-symbolic architectures \cite{Bommasani2021,OpenAI2023}. Regardless of the paradigm, AI pipelines follow a similar lifecycle:
\begin{enumerate}
    \item \textbf{Data collection and curation}, which raises issues of privacy, intellectual property, and representational fairness \cite{Bender2021};
    \item \textbf{Model training and alignment}, where biases, harmful knowledge, and unbalanced datasets can propagate societal risks \cite{Weidinger2022};
    \item \textbf{Deployment and inference}, where hallucination, misuse, and opaque decision-making can harm users at scale \cite{Ji2023};
    \item \textbf{Post-deployment monitoring}, where drift detection, auditing, and accountability mechanisms remain underdeveloped \cite{Mitchell2019}.
\end{enumerate}

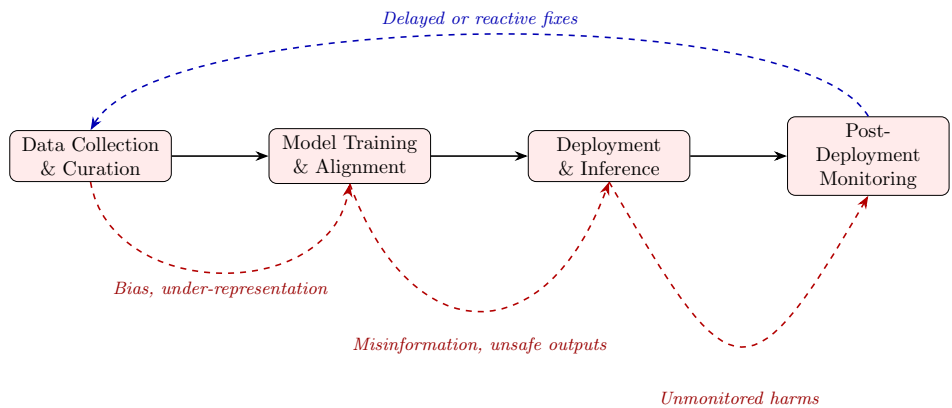
\begin{figure}[H]
\centering
\scalebox{0.75}{ 
\begin{tikzpicture}[
  stage/.style={rectangle, draw, rounded corners, align=center,
                minimum height=8mm,
                text width=26mm,
                fill=red!8},
  arrow/.style={-{Stealth}, thick},
  risk/.style={font=\small\itshape, text=red!60!black}
]
\node[stage]                         (data)    {Data Collection\\\& Curation};
\node[stage, right=1.7cm of data]    (train)   {Model Training\\\& Alignment};
\node[stage, right=1.7cm of train]   (deploy)  {Deployment\\\& Inference};
\node[stage, right=1.7cm of deploy]  (monitor) {Post-Deployment\\Monitoring};

\draw[arrow] (data.east) -- ++(0.3cm,0) |- (train.west);
\draw[arrow] (train.east) -- ++(0.3cm,0) |- (deploy.west);
\draw[arrow] (deploy.east) -- ++(0.3cm,0) |- (monitor.west);

\draw[arrow, dashed, red!70!black]
  (data.south) to[out=-80, in=-100, looseness=1.2]
  node[risk, below, yshift=-1pt] {Bias, under-representation}
  (train.south);

\draw[arrow, dashed, red!70!black]
  (train.south) to[out=-70, in=-110, looseness=1.8]
  node[risk, below, yshift=-10pt] {Misinformation, unsafe outputs}
  (deploy.south);

\draw[arrow, dashed, red!70!black]
  (deploy.south) to[out=-60, in=-120, looseness=2.4]
  node[risk, below, yshift=-19pt] {Unmonitored harms}
  (monitor.south);

\draw[arrow, dashed, blue!70!black]
  (monitor.north) to[out=120,in=60,looseness=0.45]
  node[font=\small\itshape, text=blue!70!black, above, pos=0.5]
  {Delayed or reactive fixes} (data.north);

\end{tikzpicture}
}
\caption{How early-stage ethical failures propagate forward; monitoring feedback often arrives too late.}
\label{fig:failure-propagation}
\end{figure}

Figure \ref{fig:failure-propagation} illustrates how ethical vulnerabilities can cascade across the AI development lifecycle. Issues originating during data collection and curation, such as biased sampling or under-representation, tend to persist and amplify during model training and alignment, manifesting as misinformation or unsafe outputs during deployment and inference. Weaknesses in post-deployment monitoring exacerbate the problem, as feedback loops are often delayed or reactive rather than preventative. This sequential propagation means that late-stage interventions, while necessary, are inherently less effective than proactive measures embedded at earlier stages of the pipeline.

Traditional AI ethics guidelines emphasize fairness, accountability, and transparency, yet they rarely translate into concrete engineering practices for model pipelines \cite{Floridi2018}. Current NLP evaluation metrics (BLEU, ROUGE, Recall@k) overlook ethical dimensions, and post-hoc mitigation often lags behind real-world harms.

This paper makes three contributions. First, it proposes a control architecture that systematizes existing ethics and governance frameworks into an end-to-end operational pattern usable today and adaptable to emerging AI paradigms. Second, it formalizes the enforcement of consequentialist, deontological, and virtue-ethical commitments via stage-specific controls implemented as a triple gate (metric, governance, eco) with measurable trigger conditions, escalation paths, and auditable artefacts, and provides a crosswalk to the EU AI Act, IEEE 7000 series, and the NIST AI RMF to support conformity assessment and risk management. Third, it specifies a preregistered evaluation protocol and illustrates the framework's application to large-language-model pipelines, showing how the control structure extends to multimodal, agentic, and neuro-symbolic systems.

This work contributes a control layer that existing frameworks lack: a triple-gate at each pipeline stage with explicit, testable triggers that can be automated in CI/CD and enforced at runtime. Unlike prior lifecycle ethics approaches that emphasize principles, documentation, or post-hoc reporting, the contribution here is operational: (i) measurable thresholds that decide promotion/rollback within MLOps, (ii) governance artefacts aligned to conformity assessment, and (iii) an Eco gate that makes carbon and water budgets first-class criteria alongside accuracy and fairness. This turns ethics from an after-action audit into a gatekeeping mechanism embedded in build, deploy, and serving paths.

To position this contribution relative to existing research, this paper does not propose new ethical principles or normative theories—the consequentialist, deontological, and virtue-ethical foundations are well-established in moral philosophy—nor does it introduce new fairness metrics or evaluation methods, as the metrics referenced (demographic parity, equalized odds, KL divergence, etc.) are drawn from existing literature. Rather, the contribution is a missing abstraction layer between high-level ethical principles and operational MLOps pipelines: a formalized control architecture that enables existing principles and metrics to function as enforceable lifecycle constraints. This is analogous to how safety-critical engineering disciplines—from aviation systems to medical devices to DevSecOps practices—have developed control topologies (circuit breakers, fail-safe mechanisms, continuous integration gates) that translate design principles into operational safeguards. The framework does not claim immediate empirical superiority over existing approaches; empirical validation is intentionally separated from the design and protocol specification (see Section~\ref{sec:methodological-commitment}) to enable falsifiable evaluation without post-hoc threshold tuning.

\section{Background and AI Pipeline Overview}

Understanding the ethical dimensions of AI requires first unpacking how modern AI systems are built, deployed, and maintained. The notion of an \textit{AI pipeline} captures the multi-stage process through which raw data is transformed into operational models that interact with users and society. Each stage presents distinct ethical stakes, and lapses early in the pipeline often propagate downstream in ways that are difficult to reverse.

\subsection{From Classical AI to Modern Pipelines}

Early AI workflows were relatively linear and interpretable. Symbolic reasoning systems and small-scale statistical models were trained on curated datasets, which constrained both their capabilities and their ethical surface area \cite{Floridi2018}. Errors were easier to trace, and risks were mostly limited to system reliability or local data misuse.

Modern AI, by contrast, operates in sprawling, data-intensive pipelines. Contemporary systems exhibit three defining trends:

\begin{itemize}
    \item \textbf{Massive and heterogeneous datasets,} often scraped from the open web or user-generated platforms, which blur the boundaries between public and private information \cite{Bommasani2021,Bender2021};
    \item \textbf{Deep and opaque architectures,} such as foundation models and LLMs, capable of emergent behaviors that even their developers may not fully anticipate \cite{OpenAI2023,Ganguli2022};
    \item \textbf{Persistent and iterative lifecycles,} in which models are continuously fine-tuned, redeployed, and monitored rather than released as static products \cite{Mitchell2019}.
\end{itemize}

This evolution amplifies the ethical stakes: failures in early stages can cascade into systemic harm, and reactive interventions are often inadequate. Viewing AI development through a pipeline lens is thus essential for designing proactive, ethics-by-design interventions.

\subsection{Data Collection and Curation}

Data is the substrate of modern AI, and ethical risk begins at the point of collection. Large Language Models (LLMs) and multimodal systems ingest vast quantities of text, images, code, and audio, often without explicit user consent \cite{Bommasani2021}. Three interrelated concerns dominate this stage.

First, \textbf{privacy and consent} are easily violated by indiscriminate web scraping or by incorporating user-generated data without transparent notice. Even nominally anonymized datasets may retain identifiable traces, especially for high-dimensional media like geolocated images or conversational logs. Second, \textbf{intellectual property and creative labor} are frequently implicated: generative models may train on copyrighted material or artists’ work without attribution or compensation, raising questions of both legality and fairness. Third, \textbf{representation and bias} emerge as structural risks. Web-derived corpora tend to overrepresent English and Western cultural content, underrepresenting minority languages and perspectives. The result is a pipeline that can encode and reproduce global inequities \cite{Bender2021}.

Additionally, scholars highlight the phenomenon of \textit{data colonialism}, in which cultural and linguistic resources from the Global South are extracted for AI development without local governance or reciprocal benefit \cite{Crawford2021}. Because biases and omissions at this stage propagate forward, post-hoc corrections are costly and often incomplete.

\subsection{Model Training and Alignment}

Training is where data becomes capability. Modern foundation models are typically developed in three phases: large-scale \textbf{pretraining} on general-purpose corpora, \textbf{fine-tuning} on curated or task-specific data, and \textbf{alignment} using methods such as Reinforcement Learning with Human Feedback (RLHF) \cite{OpenAI2023}. 

Ethical risks here are multifaceted. Models inherit and often \textbf{amplify biases} present in their training data, manifesting as discriminatory predictions or skewed text generation \cite{Weidinger2022}. \textbf{Emergent capabilities} introduce further uncertainty: systems can produce plausible misinformation, offensive content, or exploitable code without explicit instruction \cite{Ganguli2022}. The \textbf{environmental footprint} of training is nontrivial; frontier LLMs require immense computational resources, contributing to carbon emissions and raising questions about the social trade-offs of scaling \cite{Strubell2020}.

Alignment attempts to steer model behavior toward socially acceptable outputs, yet it is inherently normative. RLHF encodes the values of annotators and organizations, which may not reflect diverse cultural perspectives. Overreliance on alignment without addressing upstream data bias risks producing a system that \textit{appears} safe while retaining latent ethical failures.

\subsection{Environmental Accountability in AI Pipelines}

Beyond ethical risks tied to bias, privacy, and misuse, modern AI pipelines carry substantial \textbf{environmental costs}. Training state-of-the-art models—particularly large neural architectures—demands extensive computation, which translates directly into energy use and associated carbon emissions \cite{Strubell2020,Schwartz2020,Patterson2021}. Even well-optimized training runs can consume megawatt-hours of electricity, with additional hidden costs in data storage, cooling, and hardware manufacturing \cite{Henderson2020}.

These impacts are not confined to the training stage. Data collection and preprocessing incur energy and water usage for storage, cleaning, and indexing. Inference at scale, especially for latency-sensitive applications like conversational AI, requires energy-hungry infrastructure that may operate continuously across global data centers. The environmental footprint thus extends \textit{across} the pipeline: from initial dataset curation to deployment and post-release monitoring.

While some industry leaders have begun disclosing carbon metrics for training runs \cite{Patterson2021}, systematic reporting remains rare, and standards for measurement and accountability are still evolving \cite{Henderson2020,Schwartz2020}. This lack of transparency hinders both public understanding and organizational decision-making around environmental trade-offs. Without explicit budgetary and policy constraints, the drive for larger models risks locking AI development into an unsustainable trajectory.

Embedding \textbf{environmental accountability} into AI pipelines—through stage-specific targets for CO$_2$e, water usage, and energy efficiency—ensures that sustainability is treated as a first-class governance priority. As later sections discuss, the proposed \emph{Eco gate} operationalizes this principle by making environmental compliance as non-negotiable as fairness, accuracy, and legal obligations.

\subsection{Deployment and Inference}

Once models leave the laboratory, they enter the sociotechnical domain where ethical impacts are most visible. Deployment exposes models to real users through APIs, embedded applications, or search and recommendation pipelines. At this stage, four key risks dominate:

\begin{itemize}
    \item \textbf{Hallucination and misinformation,} where LLMs generate coherent but false outputs that can mislead users or erode public trust \cite{Ji2023};
    \item \textbf{Misuse and dual-use,} as open-access generative models are repurposed for disinformation campaigns, social engineering, or automated cyberattacks;
    \item \textbf{Opacity and overreliance,} since users may not understand the limitations of a model’s reasoning process and thus defer to it uncritically;
    \item \textbf{Downstream societal effects,} such as the shaping of public discourse, knowledge access, and decision-making in IR contexts.
\end{itemize}

The next generation of agentic or tool-using AI systems will expand this risk surface. When models are empowered to take autonomous actions, ethical lapses may result in immediate and tangible harm.

\subsection{Post-Deployment Monitoring}

A future-proof ethical framework recognizes that obligations do not end at release. Models interact with evolving environments, and both technical performance and social impact can drift over time \cite{Mitchell2019}. Effective monitoring includes:

\begin{itemize}
    \item \textbf{Auditing and logging} to capture anomalous behavior and provide forensic accountability;
    \item \textbf{Bias and harm tracking} to measure real-world disparities across demographic or linguistic groups;
    \item \textbf{Responsive remediation} to update, retrain, or even suspend systems when significant risks emerge.
\end{itemize}

Today, most organizations remain reactive rather than proactive in this stage, investing heavily in model release but underinvesting in lifecycle governance. As AI becomes more persistent and autonomous, continuous monitoring and rapid response pipelines will become non-negotiable components of ethical practice.

\subsection{Conceptual Implications for NLP and IR}

Framing ethics in a pipeline context has direct implications for NLP and IR. Retrieval-Augmented Generation (RAG) systems combine corpus retrieval with LLM generation, inheriting risks from both components: unvetted data sources may lead to misleading citations, while hallucination undermines factual reliability \cite{Lewis2020}. Conversational agents illustrate the tension between user engagement, transparency, and safety. Search and recommendation engines exemplify how seemingly neutral ranking choices can shape collective knowledge and public opinion.

By reconceptualizing AI ethics as a lifecycle problem, this pipeline-based perspective lays the foundation for the philosophy-to-practice bridge that the remainder of this paper develops.

\section{Philosophical Foundations of Ethical AI}

A robust ethical AI framework must rest on principles that are both historically grounded and actionable in contemporary sociotechnical contexts. Philosophical traditions provide enduring lenses for evaluating moral responsibility, while recent AI ethics research interprets these traditions in the context of model design, deployment, and governance \cite{Mittelstadt2019,Jobin2019}.

\subsection{Consequentialism and the Logic of Outcomes}

Consequentialism, most famously articulated by Jeremy Bentham and John Stuart Mill \cite{Mill1863}, judges the morality of actions by their outcomes. In AI pipelines, this translates to a focus on the \textit{real-world impact} of models, regardless of developer intentions. Harm reduction frameworks in AI ethics are inherently consequentialist: the objective is to anticipate and minimize risks such as misinformation, systemic bias, or security misuse \cite{Weidinger2022}. 

Contemporary applications include algorithmic impact assessments, which explicitly model the potential societal outcomes of AI deployments \cite{Reisman2018}. In NLP and IR, this perspective motivates evaluation protocols that measure not only accuracy but also \textit{downstream social effects}, such as the spread of harmful stereotypes or unequal access to information.

\subsection{Deontology and the Primacy of Duties}

Deontology, associated with Immanuel Kant \cite{Kant1785}, emphasizes duties and rules rather than outcomes. From this view, an AI system is unethical if it violates intrinsic moral duties—such as respecting individual privacy or avoiding deception—even if no observable harm results. 

Modern AI ethics frameworks reflect this in the form of \textbf{compliance-oriented guidelines} and \textbf{rights-based approaches}, including the European Union’s AI Act and the OECD AI Principles \cite{EUAIAct,OECD2019}. Deontological reasoning underpins requirements for informed consent in dataset collection, truthfulness in system outputs, and non-discrimination in decision-support applications. 

In the context of NLP and IR pipelines, deontological principles support practices like \textit{data minimization} (collect only what is justified), \textit{truth-preserving summarization}, and \textit{transparent citation of sources}, even when shortcuts might improve performance metrics.

\subsection{Virtue Ethics and Responsible Practice}

Virtue ethics, traced to Aristotle’s \textit{Nicomachean Ethics} \cite{Aristotle350}, shifts the moral lens from actions or outcomes to the character and practices of moral agents. In the AI pipeline, the “agents” are the developers, researchers, and organizations whose choices shape system behavior. This perspective is increasingly influential in AI ethics under the banner of \textit{responsible AI} and \textit{ethics by design} \cite{Floridi2018,Whittlestone2019}.

Virtue ethics motivates long-term stewardship and reflective practice: engineers are called to cultivate traits like honesty (resisting deceptive model claims), humility (acknowledging uncertainty in capabilities), and prudence (foreseeing emergent risks). In practical terms, this perspective informs governance measures like:

\begin{itemize}
    \item Establishing cross-disciplinary ethics boards that sustain accountability beyond compliance;
    \item Investing in ongoing education for AI practitioners to recognize sociotechnical risks;
    \item Prioritizing documentation practices such as \textit{model cards} and \textit{datasheets} that embed reflection into engineering workflows \cite{Mitchell2019,Gebru2021}.
\end{itemize}

\subsection{Bridging Philosophy and Pipeline Stages}

While each tradition offers a distinct lens, they are complementary when mapped onto the AI pipeline described in Section 2. Consequentialism aligns naturally with post-deployment monitoring and auditing, where real-world impact is measurable. Deontology is most salient during data collection and model alignment, where duties of fairness, transparency, and privacy can be operationalized as explicit constraints. Virtue ethics provides a throughline across all stages, emphasizing the cultivation of responsible research cultures that resist the “move fast and break things” ethos in frontier AI.

This synthesis lays the foundation for an \textit{ethics-by-design} approach: pipeline interventions are not ad hoc patches but principled actions embedded from inception to monitoring, informed by enduring moral philosophy and reinforced by contemporary AI ethics scholarship. Figure~\ref{fig:philosophy-pipeline-matrix} visualizes this mapping between 
ethical traditions and AI pipeline stages as a two-dimensional matrix. The rows represent the three philosophical lenses—Consequentialism, Deontology, and Virtue Ethics,while 
the columns follow the four lifecycle stages from data collection to post-deployment monitoring.  
Each cell lists concrete engineering interventions that operationalize the corresponding philosophical 
principle at that stage, highlighting where different moral frameworks converge or diverge in practice.

\newcommand{\PhiloW}{3.5cm}   
\newcommand{\CellW}{2.5cm}    

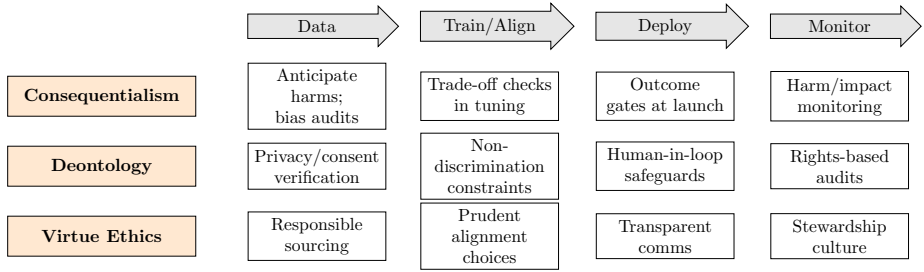
\begin{figure}[h]
\centering
\resizebox{.92\columnwidth}{!}{%
\begin{tikzpicture}[
  colsep/.style = {xshift=#1},
  arrowstage/.style = {
    draw, fill=gray!20, minimum height=8mm,
    text width=\CellW, font=\normalsize, align=center,
    single arrow, single arrow head extend=1.6mm
  },
  philosophy/.style ={rectangle, draw, align=center,
                      minimum height=8mm, text width=\PhiloW,
                      font=\bfseries\normalsize, fill=orange!18},
  cell/.style       ={rectangle, draw, align=center,
                      minimum height=8mm, text width=\CellW,
                      font=\normalsize, fill=white}
]
\node[draw=none, fill=none, minimum height=8mm, text width=\PhiloW] (hdr0) at (0,0) {};

\node[arrowstage, colsep=2.4cm] (hdr1) at (hdr0.east) {Data};
\node[arrowstage, colsep=1.8cm] (hdr2) at (hdr1.east) {Train/Align};
\node[arrowstage, colsep=1.8cm] (hdr3) at (hdr2.east) {Deploy};
\node[arrowstage, colsep=1.8cm] (hdr4) at (hdr3.east) {Monitor};

\node[philosophy] (r1) at (0,-1.4) {Consequentialism};
\node[philosophy] (r2) at (0,-2.8) {Deontology};
\node[philosophy] (r3) at (0,-4.2) {Virtue Ethics};

\node[cell] at (hdr1|-r1) {Anticipate harms;\\bias audits};
\node[cell] at (hdr2|-r1) {Trade-off checks\\in tuning};
\node[cell] at (hdr3|-r1) {Outcome\\gates at launch};
\node[cell] at (hdr4|-r1) {Harm/impact\\monitoring};

\node[cell] at (hdr1|-r2) {Privacy/consent\\verification};
\node[cell] at (hdr2|-r2) {Non-discrimination\\constraints};
\node[cell] at (hdr3|-r2) {Human-in-loop\\safeguards};
\node[cell] at (hdr4|-r2) {Rights-based\\audits};

\node[cell] at (hdr1|-r3) {Responsible\\sourcing};
\node[cell] at (hdr2|-r3) {Prudent alignment\\choices};
\node[cell] at (hdr3|-r3) {Transparent\\comms};
\node[cell] at (hdr4|-r3) {Stewardship\\culture};
\end{tikzpicture}%
}
\caption{Philosophy–practice matrix with compact spacing and larger font size.}
\label{fig:philosophy-pipeline-matrix}
\end{figure}

\section{Existing AI Ethics Frameworks and Gaps}

A variety of AI ethics frameworks and governance instruments have emerged in recent years, each reflecting different regulatory philosophies and levels of enforceability. Understanding their background is essential before evaluating their gaps in addressing full AI pipelines.

\subsection{Research-Led Ethical AI Pipeline Frameworks}

While much attention has focused on governance guidance from regulatory bodies and standards organizations, a parallel body of scholarly work has proposed ethics-by-design frameworks emerging in the post-pandemic era. These models tend to emphasise operational integration and contextual adaptability rather than solely high-level principles.

For example, the \textit{Hourglass Model of Organizational AI Governance} \cite{Mantymaki2022} introduces a multi-layered structure linking organizational policies with system-level design and lifecycle operations, enabling governance requirements to cascade through each stage of development. In the research context, the \textit{ETHICAL} framework \cite{Eacersall2024} provides actionable prompts—such as ``Examine policies'' and ``Think about social impacts''—to guide responsible use of generative AI in academic workflows. Complementing these, Al Harbi et~al.\ present \textit{Responsible Design Patterns for Machine Learning Pipelines} \cite{AlHarbi2023}, a library of reusable design components that embed ethical checks directly into technical workflows, facilitating adoption across domains.

These academic frameworks share common ground with regulatory approaches but tend to foreground researcher agency, domain-specific tailoring, and iterative ethics integration. The triple-gate model developed in this paper draws on such scholarship while extending it with an explicit environmental \emph{Eco gate} to ensure sustainability constraints are embedded alongside fairness, accuracy, and governance checks.

\subsection{Global Policy and Regulatory Frameworks}

\subsubsection{European Union AI Act.}
The EU AI Act (2021) \cite{EUAIAct} represents the first comprehensive attempt at risk-based regulation for AI. 
It classifies AI systems into four tiers:
\begin{itemize}
    \item \textbf{Unacceptable risk:} Prohibited applications such as social scoring and manipulative AI for vulnerable populations.
    \item \textbf{High risk:} Requires conformity assessments, human oversight, and detailed documentation (e.g., medical AI, recruitment tools).
    \item \textbf{Limited risk:} Must provide user disclosures when AI is in use (e.g., chatbots, deepfakes).
    \item \textbf{Minimal risk:} Permitted with no specific obligations (e.g., AI-powered spam filters).
\end{itemize}

Its pipeline implications are clearest at deployment and post-deployment stages, where documentation and human oversight are mandated. However, the Act is less prescriptive about upstream practices such as data curation and bias mitigation, leaving room for interpretation and variability across member states.

\subsubsection{NIST AI Risk Management Framework.}
The NIST AI RMF 1.0 (2023) \cite{NIST2023} reflects a U.S. approach that favors voluntary standards and lifecycle integration. 
The framework organizes governance around four functional pillars: \textit{map, measure, manage, and govern}. 
Unlike the EU’s regulatory stance, NIST RMF directly references model development pipelines, encouraging risk awareness during data collection, training, deployment, and monitoring. 

Its “govern” pillar supports organizational accountability, while “measure” and “manage” encourage proactive auditing of bias and robustness. 
Its non-binding nature, however, means that adoption varies widely across industry sectors.

\subsection{Industry Standards and Voluntary Commitments}

\subsubsection{IEEE 7000 Series and Standards-Based Ethics.}
The IEEE 7000 series \cite{IEEE7000} codifies engineering practices for value-based system design. 
IEEE 7000-2021 provides a structured process for:
\begin{enumerate}
    \item Identifying stakeholders and their value concerns;
    \item Translating values into technical system requirements;
    \item Iteratively validating that system behavior aligns with declared ethical principles.
\end{enumerate}

Table~\ref{tab:framework-comparison} summarizes these four frameworks across their lifecycle coverage, 
regulatory strength, and primary philosophical orientation. 
This comparison highlights their differing emphases and helps identify where 
pipeline stages remain under-addressed.

\begin{table}[h]
\centering
\footnotesize
\setlength{\tabcolsep}{3pt}
\caption{Comparison of major AI ethics frameworks.}
\label{tab:framework-comparison}

\begin{tabularx}{\linewidth}{p{2.6cm} *{4}{Y} p{2.2cm} p{2.6cm}}
\toprule
\textbf{Framework} & \textbf{Data} & \textbf{Train} & \textbf{Deploy} & \textbf{Monitor} & \textbf{Regulatory} & \textbf{Lens} \\
\midrule
EU AI Act \cite{EUAIAct}         & P & P & Y & Y & Binding   & Duty/rights \\
NIST AI RMF 1.0 \cite{NIST2023}  & Y & Y & Y & Y & Voluntary & Risk/outcomes \\
IEEE 7000 series \cite{IEEE7000} & Y & Y & P & P & Voluntary & Practice/design \\
EbD-AI \cite{BreyDainow2024}     & Y & Y & Y & Y & Voluntary & Values/lifecycle \\
\bottomrule
\end{tabularx}

\vspace{2pt}
{\footnotesize \textit{Legend:} Y = clear coverage; P = partial; – = limited/not explicit.}
\end{table}

Compared to the EU AI Act and NIST AI RMF, the IEEE 7000 series is highly developer-centric, embedding ethical deliberation within design stages rather than relying solely on regulatory enforcement or post-hoc auditing. 
EbD-AI, in contrast, operationalizes a fixed set of normative values across all lifecycle stages, offering strong conceptual coverage but—like IEEE 7000—remaining voluntary. 
Taken together, these contrasts show that while voluntary frameworks (IEEE, EbD-AI, NIST) provide rich design guidance, their lack of enforceability may limit their capacity to address systemic risks at scale compared to binding instruments like the EU AI Act.

\subsection{Observed Gaps Across Frameworks}

Despite their contributions, existing frameworks exhibit three persistent gaps:

\begin{itemize}
    \item \textbf{Pipeline incompleteness:} Coverage is uneven across stages, with greater specificity at deployment than in upstream data governance and model pretraining.
    \item \textbf{Limited cross-jurisdictional alignment:} Regulatory fragmentation between regions hinders global AI governance.
    \item \textbf{Weak operationalization:} High-level principles often lack concrete methods for auditing emergent behaviors in frontier models.
\end{itemize}

While most frameworks exhibit some degree of lifecycle coverage, their depth and enforceability vary. For example, the NIST AI RMF explicitly addresses all stages—from data collection to post-deployment monitoring—and encourages upstream risk awareness during data curation and model training. However, its voluntary nature means these provisions lack the systemic enforcement power of the EU AI Act. Conversely, the EU AI Act offers binding obligations, particularly for high-risk systems, but is more prescriptive at deployment than at earlier pipeline stages. 

Similarly, Brey and Dainow’s Ethics by Design for AI (EbD-AI) framework \cite{BreyDainow2024} stands out for translating six core values into phase-specific engineering tasks across the AI lifecycle, embedding ethics directly into standard development methodologies. Yet, EbD-AI’s preselected value set, qualitative task orientation, and voluntary adoption limit its adaptability and enforceability—especially for rapidly evolving paradigms such as multimodal or agentic AI.

These distinctions suggest that the limitation is not \textit{awareness} of the full lifecycle, but the \textit{operationalization} and \textit{enforcement} of that awareness in different governance contexts. The framework proposed here aims to combine the RMF’s lifecycle scope with the enforceability of risk-based regulation, while adding the concrete, metric-driven operational detail absent from both regulatory and voluntary approaches.

\section{Related Work}

Research on AI ethics in NLP and IR has evolved alongside the technological shifts from symbolic AI to deep learning and, most recently, foundation models. 
This section situates the proposed ethics-by-design framework in the context of prior work spanning three domains: 
(1) AI ethics principles and guidelines, 
(2) practical interventions for NLP and IR pipelines, 
and (3) emerging studies on ethics in large-scale and multimodal AI.

\subsection{AI Ethics Principles and Guidelines}

A significant body of work has attempted to codify the ethical responsibilities of AI systems and their developers. 
Early contributions include high-level principles such as fairness, accountability, transparency, and privacy (FAT/ML) \cite{Jobin2019,Floridi2018}. 
These efforts informed global policy initiatives, including the OECD AI Principles \cite{OECD2019}, the EU AI Act \cite{EUAIAct}, and the UNESCO Recommendation on AI Ethics (2021). 
While these guidelines established a normative baseline, scholars have noted that principles alone cannot guarantee ethical AI in practice \cite{Mittelstadt2019}. 
The “principles-to-practice gap” persists because such frameworks rarely provide operational detail for data curation, model alignment, or post-deployment monitoring.

\subsection{Pipeline-Oriented Interventions}

As NLP and IR systems grew in scale and societal impact, researchers proposed interventions to embed ethics directly into model pipelines. 
Model cards \cite{Mitchell2019} and datasheets for datasets \cite{Gebru2021} introduced structured documentation to increase transparency and support downstream auditing. 
Algorithmic impact assessments (AIA) \cite{Reisman2018} and red-teaming practices \cite{Ganguli2022} aimed to anticipate harmful outcomes during and after deployment.

In IR contexts, studies examined bias-aware document ranking, fairness in recommender systems, and mitigation of hallucination in LLM-based retrieval-augmented generation (RAG) \cite{Lewis2020,Ji2023}. 
Despite these contributions, most interventions are reactive or isolated to single stages of the pipeline, highlighting the need for a unified lifecycle framework.

\subsection{Ethics in Large-Scale and Frontier AI}

Recent literature recognizes that foundation models and LLMs pose unprecedented ethical challenges due to their emergent capabilities, opacity, and potential for dual use \cite{Bommasani2021,Weidinger2022}. 
Researchers documented systemic biases, privacy risks, and environmental costs of training \cite{Bender2021,Strubell2020}, and called for multidisciplinary oversight that spans the entire AI lifecycle. 
Initiatives like NIST AI RMF \cite{NIST2023} and IEEE 7000 series \cite{IEEE7000} reflect this lifecycle approach but rely heavily on voluntary adoption.

Recent work also highlights gaps in cross-jurisdictional governance and the difficulty of auditing frontier models at scale. 
This motivates a shift toward proactive, pipeline-integrated ethics that combines philosophical rigor with engineering feasibility—a need addressed by the proposed framework.

\subsection{Ethics by Design for AI (EbD-AI)}
Brey and Dainow’s ``Ethics by Design for AI'' (EbD-AI) methodology, developed under the EU-funded SHERPA and SIENNA projects, operationalizes six core values—human agency, privacy and data governance, fairness, well-being, transparency, and accountability—into concrete design requirements and phase-specific tasks spanning specification, design, data preparation, development, and testing \cite{BreyDainow2024}. 
Unlike principle-based guidelines such as the EU AI Act or NIST AI RMF, EbD-AI embeds ethics directly into standard development methodologies via task mapping, making ethical considerations part of everyday engineering practice rather than post-hoc audits. 
While its lifecycle integration is a notable strength, EbD-AI preselects a fixed set of values, focuses on qualitative tasks rather than metricized decision thresholds, and—like other voluntary frameworks—depends on organizational uptake for enforcement.

\subsection{Operational Distinctions from Existing Frameworks}

\begin{table}[h]
\centering
\footnotesize
\caption{Operational differences between major frameworks and the proposed approach.}
\renewcommand{\arraystretch}{4} 
\setlength{\tabcolsep}{4pt}
\hspace*{-1.25cm} 
\begin{tabular}{
  P{2.3cm}  
  P{2.7cm}  
  P{2.7cm}  
  P{2.7cm}  
  P{2.7cm}  
}
\hline
\thead{Feature} &
\thead{EU AI Act} &
\thead{NIST AI RMF} &
\thead{EbD-AI} &
\thead{Proposed\\Framework} \\
\hline
\makecell[l]{Philosophy\\integration} &
\makecell[l]{Implicit\\rights-based} &
\makecell[l]{Implicit\\risk/outcomes} &
\makecell[l]{Six fixed ethics\\values mapped\\to lifecycle} &
\makecell[l]{Maps 3 traditions\\to each lifecycle\\stage} \\
\makecell[l]{Trigger\\thresholds} &
\makecell[l]{Principles only} &
\makecell[l]{Encouraged,\\unspecified} &
\makecell[l]{Value-linked; no\\quantitative\\triggers} &
\makecell[l]{Metric-driven\\triggers per stage} \\
\makecell[l]{Adaptability to\\new paradigms} &
\makecell[l]{Amendments\\required} &
\makecell[l]{Flexible,\\informal} &
\makecell[l]{Generic model\\mappable to\\any method;\\fixed values} &
\makecell[l]{Built-in\\reassessment\\triggers} \\
\makecell[l]{Implementation\\granularity} &
\makecell[l]{Regulatory\\obligations} &
\makecell[l]{Voluntary\\best practices} &
\makecell[l]{Task-based\\interventions\\in dev process} &
\makecell[l]{Stage-by-stage\\engineering\\interventions} \\
\makecell[l]{Enforcement} &
\makecell[l]{Binding for\\high-risk} &
\makecell[l]{Voluntary} &
\makecell[l]{Voluntary;\\org-dependent} &
\makecell[l]{Works in\\voluntary and\\binding compliance} \\
\hline
\end{tabular}
\label{tab:operational-comparison}
\end{table}

Table~\ref{tab:operational-comparison} compares the operational characteristics of the EU AI Act, NIST AI RMF, EbD-AI, and the proposed framework.  
Three patterns stand out. First, existing frameworks vary in how explicitly they integrate ethical philosophy—rights-based in the EU AI Act, risk/outcomes-focused in NIST, fixed values in EbD-AI—whereas the proposed framework maps three philosophical traditions directly to each lifecycle stage.  
Second, only the proposed framework defines quantitative, metric-driven triggers for interventions, enabling automated escalation in CI/CD or MLOps contexts.  
Third, while all others require significant adaptation to address paradigm shifts, the proposed framework embeds reassessment triggers by design, allowing it to operate effectively in both voluntary and regulatory compliance environments.

The framework differs from existing approaches in three operational respects:

\begin{itemize}
    \item \textbf{Lifecycle Integration with Philosophy.}  
    Existing frameworks rarely embed normative reasoning directly into engineering checkpoints. The framework links consequentialism, deontology, and virtue ethics to measurable actions at each stage, ensuring moral reasoning remains operational rather than aspirational.

    \item \textbf{Metric-Driven Triggers.}  
    Most guidelines provide high-level principles but lack concrete thresholds for intervention. The author prescribes tool-agnostic metrics (e.g., fairness differential $>$ 0.1, KL divergence $>$ 0.15) and explicit escalation triggers that can be implemented in CI/CD or MLOps pipelines.

    \item \textbf{Future-Proofing via Adaptability.}  
    While other frameworks assume current paradigms, this one is designed to accommodate emerging risk classes in multimodal, agentic, and neuro-symbolic AI without wholesale redesign, using ``reassessment triggers'' baked into governance processes.
\end{itemize}

\section{Toward a Future-Proof Ethics-by-Design Framework}

Building on the surveyed philosophical traditions and existing AI governance frameworks, the author proposes a forward-looking \textit{ethics-by-design} framework. 
This framework integrates normative principles into the AI pipeline from inception to post-deployment, aiming to bridge the persistent gap between high-level guidelines and engineering practice. 
Unlike reactive approaches that rely on post-hoc mitigation, ethics-by-design embeds moral reasoning, risk assessment, and accountability mechanisms as first-class elements of model development.

This proactive stance contrasts with the common ``retrofit'' model, where AI components are bolted onto pipelines designed without them. Retrofitted systems often lack the data structures, governance checkpoints, and feedback loops necessary for safe and effective AI operation, resulting in brittle integrations and higher ethical risk. In contrast, AI-first pipelines can align domain workflows, data governance, and model interfaces from inception, making it easier to apply the proposed ethics-by-design framework consistently across the lifecycle.

\subsection{Philosophical Lenses}

The framework integrates three traditions in moral philosophy—consequentialism, deontology, and virtue ethics—into the lifecycle of AI systems. Each is not treated as an abstract anchor but as a practical driver for engineering decisions from the outset.

\textbf{Consequentialism} evaluates actions by their outcomes, making it essential to anticipate, measure, and minimize both intended and unintended consequences of AI systems. In engineering terms, this translates into rigorous post-deployment harm audits, continuous monitoring for disparate impact, and escalation protocols when performance drifts toward harmful behaviour.

\textbf{Deontology} focuses on duties and adherence to rules regardless of the consequences. Within the AI pipeline, this becomes enforceable compliance gates during deployment, hard constraints in model alignment, and mandatory human-in-the-loop reviews for decisions in high-stakes contexts.

\textbf{Virtue ethics} emphasizes the cultivation of good judgement and moral responsibility among practitioners. Operationally, this drives a culture of cautious iteration, stakeholder engagement in design reviews, and institutional norms that reward raising ethical risks even when doing so slows delivery.

By linking each philosophical lens to concrete engineering practices, the framework ensures that ethical reasoning and technical implementation remain connected from the earliest stages of design.

Each philosophical lens constrains a distinct failure mode rather than serving as post hoc rhetoric: consequentialism bounds outcome harms via measurable risk thresholds, deontology enforces rights and procedural duties through non-negotiable prohibitions and documentation requirements, and virtue ethics shapes organizational character through oversight cadence, escalation norms, and accountability roles.

\subsection{Core Principles}

The framework rests on three mutually reinforcing principles:

\begin{itemize}
    \item \textbf{Pipeline Integration} — Ethical oversight is mapped explicitly onto data collection, model training, deployment, and post-deployment monitoring stages. Each stage has dedicated checks for privacy, fairness, and downstream societal impact.
    \item \textbf{Multi-Layered Accountability} — Responsibility is distributed across individual researchers, project teams, and organizations. This reflects virtue ethics’ focus on responsible practice, deontology’s attention to duties, and consequentialism’s emphasis on outcomes. The framework is designed for interoperability with varying legal regimes and adaptable to non-Western governance models, enabling consistent ethical oversight across jurisdictions with differing regulatory maturity and cultural norms.
    \item \textbf{Dynamic Adaptation} — As AI capabilities evolve—toward multimodal, agentic, or neuro-symbolic systems—the framework supports iterative reassessment of risks and principles, ensuring that oversight keeps pace with technological change. This includes reliance on tool-agnostic metrics such as fairness differentials, energy use per training hour, or embedding drift scores, and evaluation methods like bias audit pipelines, adversarial red-team simulations, and post-deployment harm incidence tracking to maintain consistent comparability across architectures and modalities.
\end{itemize}

\noindent In practical terms, ``future-proof'' in this context means three things:  
\begin{inparaenum}[(i)]
    \item the inclusion of explicit triggers for reassessment when model capabilities, usage contexts, or societal expectations change;
    \item reliance on tool-agnostic metrics—such as fairness differentials, energy use per training hour, or embedding drift scores—that remain applicable across architectures and modalities; and
    \item a governance process designed to incorporate unanticipated risk classes without requiring wholesale redesign of the framework.
\end{inparaenum}
By treating adaptability as an operational requirement rather than an aspirational quality, the framework seeks to remain relevant through successive AI paradigm shifts.

\subsection{Anticipating Risks in Emerging AI Paradigms}

The framework’s future-proofing claim rests on its adaptability to paradigms whose risk profiles are not yet fully known. 
To illustrate, consider three examples:
\begin{itemize}
    \item \textbf{Multimodal Systems:} A cross-modal model may infer sensitive attributes (e.g., health status from voice, socio-economic class from home interior) without explicit user disclosure, raising novel consent and profiling concerns.
    \item \textbf{Agentic AI:} Autonomous decision-making agents coordinating across networks may trigger cascading effects—such as self-initiated transactions or negotiations—that lack human-in-the-loop oversight at critical junctures.
    \item \textbf{Neuro-symbolic Models:} Symbolic reasoning components can produce discrete, unambiguous outputs that appear authoritative but embed unexamined logical biases; their opacity may differ from deep nets but still hinder auditability.
\end{itemize}
These scenarios illustrate that “future-proof” cannot mean “unchanging”; rather, it implies the framework’s capacity for iterative reassessment as new risk classes emerge.

\subsection{Methodological Commitment and Preregistration}
\label{sec:methodological-commitment}

This paper contributes a design specification and evaluation protocol rather than empirical results from executed experiments. This separation is intentional and methodologically motivated. Empirical evaluation of gate-based ethics frameworks faces a fundamental challenge: threshold selection and metric calibration are inherently domain- and context-dependent, and post-hoc adjustment of thresholds after observing outcomes introduces selection bias and undermines scientific rigor.

To address this, the framework is accompanied by a preregistered evaluation protocol (Appendix~\ref{app:prereg}) that specifies endpoints, sampling procedures, labeling criteria, and analysis plans \textit{before} empirical execution. This preregistration serves multiple purposes: (i) it ensures falsifiability by defining success criteria ex ante, (ii) it prevents post-hoc threshold tuning that could inflate apparent effectiveness, (iii) it enables reproducible evaluation across different organizations and domains, and (iv) it makes explicit the trade-offs between false-positive rates (blocking safe deployments) and false-negative rates (allowing risky deployments) that any gate-based system must navigate.

The protocol specifies primary endpoints (gate false-positive and false-negative rates by stage and metric), secondary endpoints (risk reduction, time-to-detection, emissions–quality–latency trade-offs), and sensitivity analyses across predefined threshold bands. By committing to these specifications before execution, the framework can be evaluated without the circularity that would arise from tuning thresholds to match observed outcomes. This methodological commitment positions the work as a falsifiable design contribution rather than a post-hoc case study.

\subsection{Operationalizing the Framework}

This section first presents an end-to-end illustration of the triple-gate across the lifecycle, then details a reference implementation (Section~\ref{sec:ref-impl}), and finally points to the preregistered evaluation protocol (Appendix~\ref{app:prereg}).

To translate principles into actionable engineering steps, the four-stage operational loop is expanded with concrete metrics, tools, and illustrative examples. Each stage is treated as a continuous practice rather than a one-time checkpoint, ensuring ethics remains embedded throughout the AI lifecycle.

Although presented sequentially for clarity, these stages form a cyclical process rather than a one-way pipeline. Insights from post-deployment monitoring feed directly into updates in data governance and alignment practices; alignment decisions may prompt changes to deployment protocols; and new risks identified at any point can trigger a full reassessment starting from the data stage. This feedback loop ensures that the framework is capable of continuous self-correction as contexts and capabilities evolve.

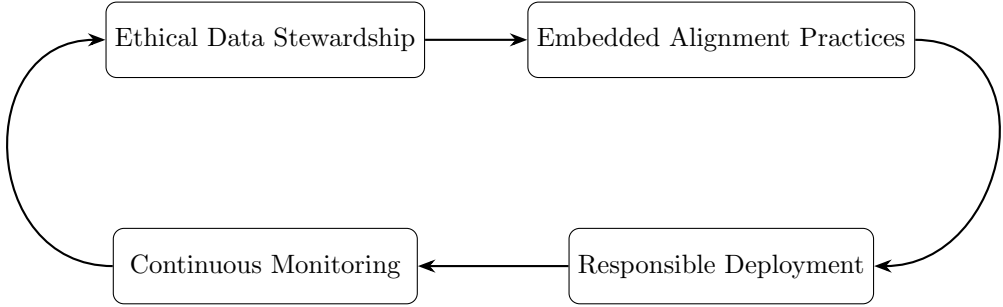
\begin{figure}[h]
\begin{tikzpicture}[xshift=-8mm, 
    stage/.style={rectangle, draw, rounded corners, align=center, minimum width=4cm, minimum height=1cm},
    arrow/.style={-{Stealth}, thick}
]

\path[use as bounding box] (-3.25,1.2) rectangle (3.25,-4.2);

\node[stage] (data) at (0,0) {Ethical Data Stewardship};
\node[stage] (align) at (6,0) {Embedded Alignment Practices};
\node[stage] (deploy) at (6,-3) {Responsible Deployment};
\node[stage] (monitor) at (0,-3) {Continuous Monitoring};

\draw[arrow] (data.east) -- (align.west);
\draw[arrow] (deploy.west) -- (monitor.east);

\draw[arrow] (align.east) to[out=0,in=0,looseness=1.5] (deploy.east);

\draw[arrow] (monitor.west) to[out=180,in=180,looseness=1.5] (data.west);

\end{tikzpicture}
\caption{Cyclical structure of the operational loop, where each stage can inform and trigger updates to earlier stages.}
\end{figure}

\subsubsection{\textbf{Stage 1:} Ethical Data Stewardship.}
Before training begins, datasets undergo structured assessment using \textit{datasheets for datasets}~\cite{Gebru2021} and algorithmic impact templates~\cite{Reisman2018}. Practitioners evaluate coverage (e.g., data coverage ratio, language representation index), verify consent, and detect over- or under-represented communities; lightweight profiling tools such as \texttt{pandas-profiling}~\cite{pandas_profiling} and \texttt{Great Expectations}~\cite{great_expectations} can automate much of this work. In parallel, sourcing and curation decisions incorporate sustainability as a first-class constraint: minimize redundant data to reduce downstream compute, prefer storage and processing in low-carbon cloud regions, and log estimated storage/transfer energy and cooling water use so that environmental cost is visible at the same decision points as fairness and quality~\cite{Strubell2020,Crawford2021}. For example, a multilingual NLP project might quantify the proportion of documents per ISO 639-1 language code and flag those falling below a 2\% threshold for augmentation \emph{while} recording the projected storage and transfer footprint, enabling a trade-off analysis between representational coverage and carbon/water budget before data growth locks in higher lifecycle costs. We use ISO 639-1 language codes as a non-PII proxy group suitable for multilingual support, and—where lawful and appropriate—layer additional groups to capture intersectional effects under privacy-preserving aggregation.

\subsubsection{\textbf{Stage 2:} Embedded Alignment Practices.}
During fine-tuning and alignment, ethical checkpoints are woven into standard evaluation workflows. Bias is measured across protected attributes using metrics such as \textit{demographic parity difference} (DPD) and \textit{equalized odds} (EO), with toolkits like IBM \texttt{AIF360}~\cite{aif360} or Microsoft \texttt{Fairlearn}~\cite{fairlearn}. In parallel, environmental performance is monitored with tools such as \texttt{CarbonTracker}~\cite{carbontracker}, CodeCarbon, or ML~CO\textsubscript{2}~Impact, which log electricity consumption, carbon emissions (kg~CO\textsubscript{2}e), and, where available, water usage during each training and evaluation cycle~\cite{Strubell2020,Crawford2021}. This ensures that model quality metrics (e.g., accuracy, fairness) are contextualized with sustainability costs, enabling trade-offs such as early stopping or low-precision training when marginal accuracy gains carry disproportionate environmental impact. For example, when aligning an LLM with RLHF, annotator pools should be selected for value diversity, bias reports generated after each epoch to guide dataset rebalancing, and training schedules adjusted to run in regions/times with higher renewable energy availability to reduce carbon intensity.

\textit{Worked example:} Consider a multilingual customer support chatbot deployed in both the U.S. and the EU. In the U.S., \textit{protected attributes} typically include race, gender, religion, and disability status under U.S. Equal Employment Opportunity Commission (EEOC) guidelines; in the EU, the General Data Protection Regulation (GDPR)’s “special categories” extend protection to attributes such as political opinions, trade union membership, and biometric identifiers. Suppose the model recommends a refund in $65\%$ of cases for Group~A and $55\%$ for Group~B. The demographic parity difference (DPD) is calculated as:
\[
\text{DPD} = |0.65 - 0.55| = 0.10
\]
If the framework threshold is set to $\text{DPD} > 0.1$, this result would be flagged for mitigation. Simultaneously, environmental logs might reveal that training this alignment phase consumed 450~kWh of electricity and 350~liters of cooling water, prompting consideration of smaller model checkpoints or quantization in the next cycle to meet sustainability thresholds.

\subsubsection{\textbf{Stage 3:} Responsible Deployment and Guardrails.}
Prior to release, pre-deployment audits stress-test the model against misuse scenarios and high-risk failure modes. These include adversarial red-teaming~\cite{Ganguli2022} to measure prompt injection success rates, automated toxicity checks via APIs such as Perspective~\cite{perspective_api}, and factual verification through retrieval-based methods. Alongside safety and accuracy, deployment readiness assessments should incorporate an environmental readiness check: projecting the carbon footprint, electricity demand, and water usage of the intended inference workload~\cite{Strubell2020,Crawford2021}. For high-throughput or latency-critical applications, strategies such as model distillation, quantization, or on-device inference can reduce operational energy use, while deploying on data centers powered by renewable energy mitigates climate impact. 

Guardrails—such as explicit refusal templates, safe-mode fallbacks, and rate-limiting for high-cost queries—are tuned based on these audits. For example, if a prompt injection success rate exceeds 1\%, guardrail triggers might be tightened and a deployment delay introduced until mitigations are validated; likewise, if projected emissions exceed the organization’s per-month CO\textsubscript{2}e budget, model optimization or capacity throttling may be enforced before go-live.

Pre-deployment serving-emissions budgets are treated as scenario-based estimates (low, expected, high) and are re-bound at the first monitoring window; from that point, the Monitor-stage Eco gate serves as the canonical arbiter that supersedes earlier estimates.

\subsubsection{\textbf{Stage 4:} Continuous Monitoring and Iterative Governance.}  
After deployment, monitoring captures both technical drift and social impact. Drift scores (e.g., KL divergence between embedding distributions over time) can signal shifts in model behaviour, while harm incidence rates and stakeholder satisfaction indices track real-world effects. Tools such as \texttt{evidently.ai} \cite{evidently_ai} can automate drift detection, while user-facing portals allow stakeholders to report harms.  

Environmental performance should also be monitored as part of the governance cycle, since inference workloads at scale can have non-trivial carbon and water footprints. Following Strubell et~al.~\cite{Strubell2020}, operators can periodically estimate CO\textsubscript{2}-equivalent emissions by logging inference GPU-hours, local grid carbon intensity, and data centre PUE (Power Usage Effectiveness). For facilities using evaporative cooling, water usage metrics—highlighted in Crawford's analysis of AI infrastructure costs \cite{Crawford2021}—should be tracked, especially in water-scarce regions.  

\textit{Worked example:} A deployed multilingual chatbot serving 20 million queries per month logs its total inference GPU-hours and calculates monthly CO\textsubscript{2} emissions using local grid emission factors. If emissions exceed a set budget (e.g., 50 kg CO\textsubscript{2}e per million queries), governance protocols require optimization actions—such as model distillation, quantization, or routing low-risk queries to smaller models—before the next reporting cycle. Similarly, water usage per query is monitored, and if cooling-related consumption surpasses regional conservation thresholds, workloads may be migrated to less water-stressed facilities.

\subsubsection{End-to-End Illustration.}  
Consider a conversational AI deployed in a multilingual customer support setting. 
During Ethical Data Stewardship, the dataset is profiled with \texttt{pandas-profiling} to confirm that each target language accounts for at least 10\% of the training corpus. 
In Embedded Alignment Practices, IBM \texttt{AIF360} computes a demographic parity difference for resolution suggestions by customer gender, flagging values exceeding 0.1 as requiring mitigation. 
Responsible Deployment includes a Perspective API toxicity check, requiring a mean score below 0.05 on a curated red-team prompt set before go-live. 
Finally, Continuous Monitoring with \texttt{evidently.ai} triggers retraining if KL divergence in response embeddings over a 30-day window exceeds 0.15, or if monthly user harm reports surpass a pre-set tolerance of five incidents.

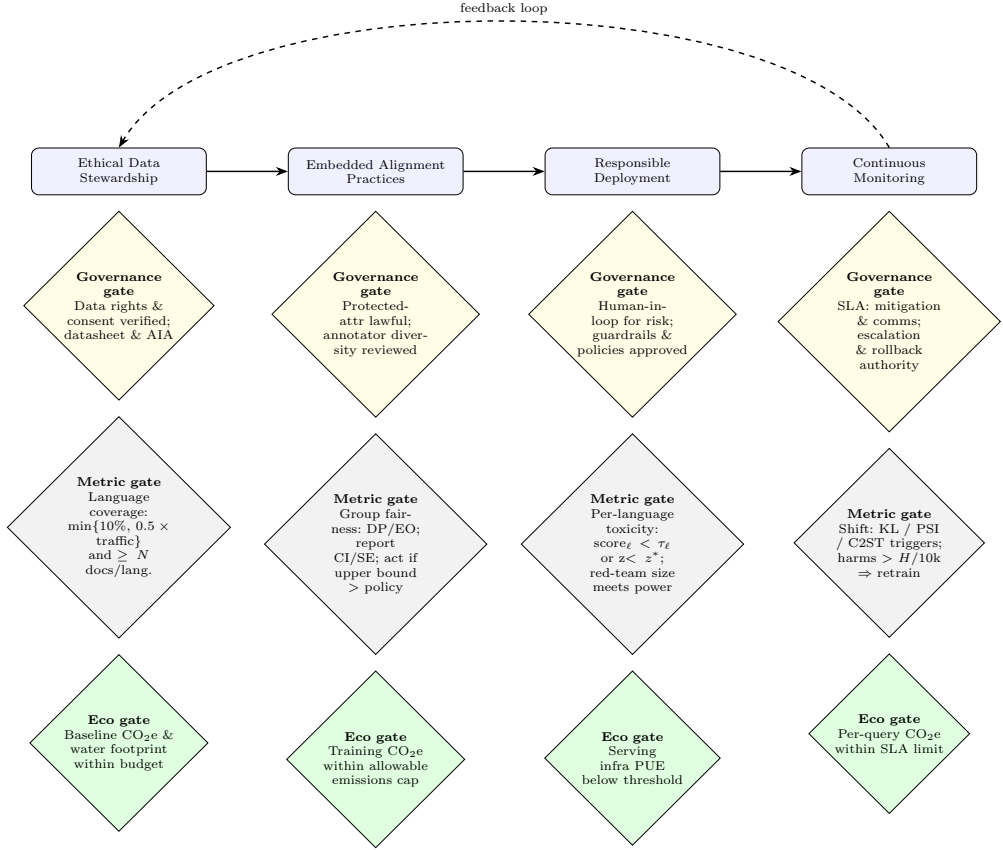
\vspace{-0.75em} 

\begin{figure}[H]
\centering
\resizebox{\columnwidth}{!}{%
\begin{tikzpicture}[
  font=\footnotesize,
  arrow/.style={-{Stealth}, thick},
  stage/.style={rectangle, rounded corners, draw, align=center,
                minimum height=9mm, text width=31mm, fill=blue!6},
  gateM/.style={diamond, draw, align=center, fill=gray!10,
                minimum width=28mm, minimum height=16mm,
                inner sep=1pt, text width=22mm},
  gateE/.style={diamond, draw, align=center, fill=green!12,
                minimum width=28mm, minimum height=16mm,
                inner sep=1pt, text width=22mm},
  gateG/.style={diamond, draw, align=center, fill=yellow!12,
                minimum width=28mm, minimum height=16mm,
                inner sep=1pt, text width=22mm}
]

\def\stageSep{1.55cm}
\def\gateV{3mm}

\node[stage] (s1) {Ethical Data Stewardship};
\node[stage, right=\stageSep of s1] (s2) {Embedded Alignment\\Practices};
\node[stage, right=\stageSep of s2] (s3) {Responsible\\Deployment};
\node[stage, right=\stageSep of s3] (s4) {Continuous\\Monitoring};

\draw[arrow] (s1) -- (s2);
\draw[arrow] (s2) -- (s3);
\draw[arrow] (s3) -- (s4);

\draw[arrow, dashed] (s4.north) to[out=120,in=60,looseness=0.65]
  node[above]{feedback loop} (s1.north);


\node[gateG, below=\gateV of s1] (g1) {\textbf{Governance gate}\\Data rights \& consent verified;\\datasheet \& AIA};
\node[gateM, below=\gateV of g1] (m1) {\textbf{Metric gate}\\Language coverage:\\$\min\{10\%,\,0.5\times\text{traffic}\}$\\and $\ge N$ docs/lang.};
\node[gateE, below=\gateV of m1] (e1) {\textbf{Eco gate}\\Baseline CO$_2$e \&\\water footprint within budget};

\node[gateG, below=\gateV of s2] (g2) {\textbf{Governance gate}\\Protected-attr lawful;\\annotator diversity reviewed};
\node[gateM, below=\gateV of g2] (m2) {\textbf{Metric gate}\\Group fairness: DP/EO;\\report CI/SE; act if\\upper bound $>$ policy};
\node[gateE, below=\gateV of m2] (e2) {\textbf{Eco gate}\\Training CO$_2$e\\within allowable emissions cap};

\node[gateG, below=\gateV of s3] (g3) {\textbf{Governance gate}\\Human-in-loop for risk;\\guardrails \& policies approved};
\node[gateM, below=\gateV of g3] (m3) {\textbf{Metric gate}\\Per-language toxicity:\\score$_\ell<\tau_\ell$ or z$<z^*$;\\red-team size meets power};
\node[gateE, below=\gateV of m3] (e3) {\textbf{Eco gate}\\Serving infra PUE\\below threshold};

\node[gateG, below=\gateV of s4] (g4) {\textbf{Governance gate}\\SLA: mitigation \& comms;\\escalation \& rollback authority};
\node[gateM, below=\gateV of g4] (m4) {\textbf{Metric gate}\\Shift: KL / PSI / C2ST triggers;\\harms $>H$/10k $\Rightarrow$ retrain};
\node[gateE, below=\gateV of m4] (e4) {\textbf{Eco gate}\\Per-query CO$_2$e\\within SLA limit};

\end{tikzpicture}%
}

\vspace{0.35em}
\footnotesize
\begin{minipage}{0.96\columnwidth}
\textbf{Reading the diagram.} Each stage has three gates in this order: \emph{Governance} (yellow), \emph{Metric} (gray), and \emph{Eco} (green). Eco gates enforce stage-specific environmental limits (CO$_2$e, water, energy) alongside performance and compliance criteria.
\end{minipage}

\caption{Triple-gate lifecycle for a multilingual support. Gates run as CI/CD jobs; failure blocks promotion; successes emit signed artefacts.}
\label{fig:triple-gated-pipeline}
\end{figure}

The triple-gate structure in Figure~\ref{fig:triple-gated-pipeline} illustrates how each stage of the pipeline is governed by a quantitative \emph{Metric gate}, a qualitative \emph{Governance gate}, and an environmental \emph{Eco gate}.  
At the \textbf{Ethical Data Stewardship} stage, the Metric gate enforces a language coverage target—such as a minimum of 10\% or half the observed traffic share, with an absolute document count threshold—while the Governance gate verifies data rights and consent through datasheets \cite{Gebru2021} and algorithmic impact assessments \cite{Reisman2018}. The Eco gate estimates dataset storage and preprocessing energy using tools such as \texttt{CodeCarbon} or \texttt{CarbonTracker}, reporting emissions in CO$_2$e and estimated water usage following emerging sustainable AI reporting standards \cite{Henderson2020,Schwartz2020}.

In \textbf{Embedded Alignment Practices}, the Metric gate applies group fairness tests such as demographic parity (DP) and equalized odds (EO) \cite{hardt2016equality}, reporting confidence intervals (CI) and standard errors (SE) to ensure statistical robustness, while governance reviews check that handling of protected attributes complies with applicable law and that annotator pools are demographically diverse. The Eco gate logs per-epoch GPU hours, electricity source mix, and cooling water usage, enabling carbon intensity benchmarking \cite{Patterson2021}.  

The \textbf{Responsible Deployment} Metric gate measures per-language toxicity, requiring scores in each supported language to remain below a set threshold $\tau_\ell$ or a critical z-score, with red-team prompt sets large enough to meet statistical power requirements \cite{Ganguli2022}; its Governance gate mandates human-in-the-loop sign-off for high-risk use cases and formal approval of guardrails and policies \cite{Floridi2018}. The Eco gate conducts a final pre-release life-cycle assessment, estimating cumulative CO$_2$e emissions from training to inference, and flags deployments that exceed the organization’s environmental budget for review \cite{Schwartz2020}.  

Finally, \textbf{Continuous Monitoring} uses shift-detection metrics, such as Kullback–Leibler (KL) divergence \cite{kullback1951information}, Population Stability Index (PSI) \cite{PSIguide}, and Classifier Two-Sample Test (C2ST) \cite{lopez2016revisiting}—to trigger retraining when any exceed thresholds, and tracks severity-weighted user harm rates per 10{,}000 interactions ($H$), while the Governance gate enforces service-level agreements (SLA) for mitigation, communication, and escalation. The Eco gate monitors inference energy per request, aggregated CO$_2$e over time, and water consumption in cooling, ensuring that efficiency degradations are caught alongside accuracy drops \cite{Henderson2020,Patterson2021}.  

Together, the three gates at each stage ensure that measurable performance, compliance, and sustainability criteria are met before progression, reducing the likelihood of ethical, technical, or environmental harms propagating downstream.

The illustration above specifies what the triple-gate checks and why. The next subsection translates that policy logic into a concrete implementation blueprint suitable for CI/CD and runtime serving. Readers focused on mechanics and audit artefacts can continue to Section~\ref{sec:ref-impl}; those looking for the empirical protocol can jump to Appendix~\ref{app:prereg}.

\subsubsection{Reference implementation of the triple-gate}
\label{sec:ref-impl}

Building on the end-to-end illustration, this subsubsection specifies how to integrate the triple-gate into CI/CD and serving, and what artefacts are produced for audits. It is implementation-oriented and does not report quantitative results.

Gate API contract. Each stage exposes a gate endpoint that consumes a metrics bundle and returns a signed decision with justifications and pointers to artefacts.

\begin{verbatim}
POST /gate/{stage}
Request:
  metrics: {metric_name: value, ...}
  metadata: {commit_id, model_id, timestamp, actor}
Response:
  decision: pass|block|throttle
  reasons: [ {metric, value, threshold, rule_id}, ... ]
  artefacts: [ {path, sha256, signer_id}, ... ]
  audit_id: UUID
\end{verbatim}

CI/CD hook. The gates run as blocking jobs; failures prevent promotion and emit auditable logs.

\begin{verbatim}
stages: [build, test, gates, deploy]
jobs:
  - name: gate:data
    run: gates.py --stage data --config cfg.yaml --out artefacts/data/
    on_fail: block
  - name: gate:train_align
    run: gates.py --stage train_align --config cfg.yaml --out artefacts/train/
    on_fail: block
  - name: gate:deploy
    run: gates.py --stage deploy --config cfg.yaml --out artefacts/deploy/
    on_fail: block
  - name: gate:monitor
    run: gates.py --stage monitor --config cfg.yaml --out artefacts/monitor/
    on_fail: alert
\end{verbatim}

Artefact manifest. Gate runs emit a compact manifest used for conformity assessment and audits.

\begin{verbatim}
audit_id: UUID
stage: data|train_align|deploy|monitor
entries:
  - gate: metric|governance|eco
    metric: name
    value: float|bool
    threshold: float|bool
    decision: pass|block|throttle
    evidence: path/to/file
    timestamp: ISO-8601
    signer: key-id
\end{verbatim}

\paragraph{Threat-scenario walkthroughs}
Three scenarios illustrate the trigger \textrightarrow{} automated action \textrightarrow{} oversight pattern. No empirical claims are made here; thresholds are defined in the configuration and may vary by domain.

\begin{description}
  \item[Scenario A — Data fairness regression.] 
  Trigger: demographic parity difference exceeds the configured threshold for a language group. 
  Action: block promotion and open a remediation ticket with the offending slice and seed examples. 
  Oversight: responsible AI reviewer approves release only after a fresh gate pass.

  \item[Scenario B — Post-alignment toxicity spike.] 
  Trigger: toxicity mean breaches the stage threshold or misuse block-rate falls below the minimum. 
  Action: block and roll back to the last green model snapshot, then re-run alignment tests. 
  Oversight: safety lead signs off on the mitigation and re-test report.

  \item[Scenario C — Eco budget breach at deploy.] 
  Trigger: projected serving emissions per query exceeds the regional budget. 
  Action: throttle or block; propose quantization or routing to a smaller model. 
  Oversight: platform owner records trade-offs and signs the release record.
\end{description}

These walkthroughs operationalize the triple-gate as enforceable controls with auditable artefacts. This concludes the reference implementation; the preregistered empirical protocol is specified in Appendix~\ref{app:prereg}.

\vspace{1em}

All thresholds are pre-registered (Appendix~\ref{app:prereg}) and sensitivity-tested within predefined bands, and governance-gate reviews discourage metric gaming and require documented justification for any post hoc change.

\subsection{Adaptation Pathways for Cross-Context Deployment}
\label{sec:adaptation-pathways}

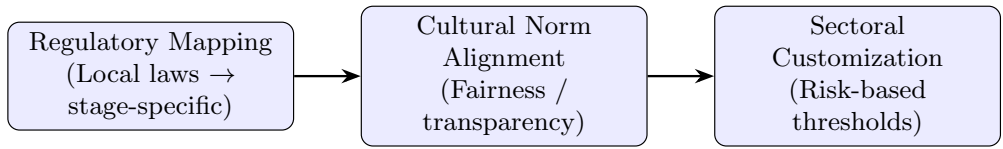
\begin{figure}[h]
\centering
\resizebox{\columnwidth}{!}{%
\begin{tikzpicture}[
  node/.style={rectangle, draw, rounded corners, align=center,
               fill=blue!8, text width=3.2cm, minimum height=1cm,
               font=\small}
]
\node[node] (reg) {Regulatory Mapping\\(Local laws $\rightarrow$ stage-specific)};
\node[node, right=0.8cm of reg] (culture) {Cultural Norm Alignment\\(Fairness / transparency)};
\node[node, right=0.8cm of culture] (sector) {Sectoral\\Customization\\(Risk-based thresholds)};

\draw[-{Stealth}, thick] (reg.east) -- (culture.west);
\draw[-{Stealth}, thick] (culture.east) -- (sector.west);
\end{tikzpicture}%
}
\caption{Adaptation pathways for deploying the framework across jurisdictions and sectors.}
\label{fig:adaptation-pathways}
\end{figure}

A ``future-proof'' framework must not only survive paradigm shifts but also adapt across legal regimes, sectors, and cultural contexts.  
The author proposes three adaptation pathways (Figure~\ref{fig:adaptation-pathways}):

\begin{enumerate}
    \item \textbf{Regulatory Mapping:} Align each framework stage with local legal instruments (e.g., data minimization under GDPR, algorithmic transparency under the U.S. Algorithmic Accountability Act) while maintaining the same internal checkpoints.
    \item \textbf{Cultural Norm Alignment:} Adjust fairness and transparency metrics to reflect local normative expectations (e.g., group fairness definitions in collectivist societies vs. individual fairness in individualist societies).
    \item \textbf{Sectoral Customization:} Tailor metric thresholds and audit frequency to domain risk profiles (e.g., higher scrutiny for healthcare AI than for recommendation engines).
\end{enumerate}

These pathways allow for local compliance without sacrificing the framework’s internal coherence, ensuring that adaptations remain systematic rather than ad hoc.

\subsection{Regulatory crosswalk: mapping gates to EU AI Act and NIST AI RMF}

As summarized in Table~\ref{tab:reg-crosswalk}, the crosswalk is an illustrative mapping from the triple-gate controls to high-level obligations in the EU AI Act and to NIST AI RMF functions. The intent is alignment at the level of control purpose rather than article-by-article compliance. Governance gates map to documentation, oversight, transparency, and post-market duties; Metric gates map to measurement and pre-release safety testing; Eco gates are treated as risk-management and reporting artefacts even where the Act is silent on explicit environmental thresholds. In practice, these controls run as CI/CD gate checks whose pass or fail produces auditable artefacts for conformity assessment and ongoing risk management. Entries marked as organizational policy indicate controls that extend beyond the Act’s minimum while remaining compatible with NIST functions.

\begin{table}[h]
\centering
\caption{Crosswalk from triple-gate controls to EU AI Act obligations and NIST AI RMF functions.}
\label{tab:reg-crosswalk}
\footnotesize
\setlength{\tabcolsep}{4pt}
\begin{tabular}{p{2.4cm}p{2.1cm}p{4.2cm}p{3.8cm}}
\hline
Pipeline stage & Gate & EU AI Act (shorthand) & NIST AI RMF \\
\hline
Data         & Governance & data governance; documentation          & Map, Measure, Govern \\
Data         & Metric     & bias; data quality                      & Measure, Manage \\
Data         & Eco        & risk management; reporting              & Measure, Govern \\
Train/Align  & Governance & human oversight; records                & Govern, Manage \\
Train/Align  & Metric     & accuracy; robustness tests              & Measure, Manage \\
Train/Align  & Eco        & energy and water logs (org policy)      & Measure, Govern \\
Deploy       & Governance & transparency; human-in-the-loop         & Govern, Manage \\
Deploy       & Metric     & pre-release safety tests                & Measure, Manage \\
Deploy       & Eco        & serving budgets (org policy)            & Measure, Manage \\
Monitor      & Governance & post-market monitoring; incidents       & Govern, Manage \\
Monitor      & Metric     & drift and harm triggers                 & Measure, Manage \\
Monitor      & Eco        & footprint tracking (org policy)         & Measure, Govern \\
\hline
\end{tabular}
\end{table}

\section{Discussion}
In this research, the Eco gate is positioned a first-class blocker: deployments may pass accuracy and fairness yet still be halted if projected emissions or water use exceed the organizational budget. Treating sustainability as a gating criterion, rather than a post-hoc report, rebalances incentives for model size, quantization, and placement.

Embedding gates in CI/CD and runtime serving paths is the crucial shift from "principles and documentation" to "operational control." The gate outputs decide promotion, rollback, or throttling, and their logs become audit artefacts for conformity assessment. This is where the framework substantively departs from prior lifecycle ethics work.

The distinction between an organizational framing and a control architecture is fundamental. Many ethical failures in AI systems are not failures of principle—the principles are often well-understood—but failures of \textit{control}: the absence of enforceable checkpoints that prevent harmful configurations from progressing through the pipeline. This parallels safety-critical engineering domains: aviation safety depends not only on design principles but on control topologies (redundant systems, automated fail-safes, mandatory pre-flight checks) that enforce those principles operationally. Similarly, DevSecOps practices embed security gates into CI/CD pipelines not as documentation exercises but as blocking controls that prevent vulnerable code from reaching production.

The triple-gate framework positions ethics governance as a control topology rather than a checklist. Each gate functions as a circuit breaker: when thresholds are breached, the system blocks progression and requires remediation before resumption. This is qualitatively different from post-hoc reporting or voluntary best practices, which rely on organizational discipline rather than automated enforcement. The proposed framework differs from prior lifecycle ethics approaches in its deliberate integration of normative reasoning with concrete engineering checkpoints from project inception. While frameworks such as the NIST AI RMF or IEEE 7000 series recognize lifecycle coverage, their voluntary nature and abstract treatment of philosophical principles often leave operational details underspecified. By contrast, this approach treats philosophy not as a preamble but as an ongoing operational constraint, with metrics and governance triggers woven into each stage as enforceable controls.

This philosophy-to-practice orientation has implications for how AI is integrated into domain workflows. In the ``retrofit'' model—where AI components are appended to pipelines designed without them—core data structures, governance hooks, and feedback loops are often absent. This absence makes alignment retrofits brittle, increases risk, and constrains ethical guardrails. In contrast, AI-first pipelines can align domain processes, data governance, and model interfaces from inception, making it easier to apply lifecycle ethics consistently.

Adoption will face practical challenges. Organizational inertia may resist structural changes to pipelines. Costs associated with implementing continuous monitoring and bias audits may deter smaller organizations. Cross-jurisdictional applicability demands adaptation to non-Western governance norms, local regulatory regimes, and culturally specific conceptions of fairness and accountability. Addressing these challenges will require not only technical tooling but also policy incentives and sector-specific adaptation guidelines.

By embedding Eco gates throughout the pipeline, environmental impact assessment shifts from an afterthought to a first-class design requirement. This operationalizes recommendations from prior work \cite{Henderson2020,Schwartz2020,Patterson2021} by making carbon and energy tracking mandatory at each stage—e.g., dataset collection, model training, serving, and post-deployment monitoring. The Eco gate enables actionable trade-off analysis: a deployment might pass accuracy and fairness tests but be halted if projected emissions exceed organizational or policy thresholds. In doing so, the framework reframes sustainability not as a constraint that slows innovation, but as a proactive criterion that can guide model architecture, hardware choice, and deployment strategy.

\subsection{Limitations}

Several limitations constrain the framework's applicability and should inform its adoption. First, this paper presents a design specification and preregistered evaluation protocol rather than empirical results from executed experiments. While this separation is methodologically motivated (see Section~\ref{sec:methodological-commitment}), it means that claims about effectiveness, false-positive/false-negative rates, and risk reduction remain untested until the protocol is executed. The framework's value as a control architecture is independent of empirical validation, but its practical impact depends on demonstrating that gates reduce harmful deployments without excessive false positives.

Second, threshold selection and calibration are inherently sensitive to domain, context, and organizational risk tolerance. The framework specifies that thresholds should be preregistered and sensitivity-tested, but it cannot prescribe universal values. This creates a risk of threshold gaming: organizations might set permissive thresholds to avoid blocking deployments, undermining the framework's protective function. Governance gates mitigate this by requiring documented justification for threshold choices and independent review, but the risk remains, especially in voluntary adoption contexts.

Third, the framework assumes a baseline level of organizational maturity: it requires CI/CD infrastructure, monitoring capabilities, and governance processes that may be absent in smaller organizations or early-stage projects. The costs of implementing continuous monitoring, bias audits, and environmental tracking may deter adoption, particularly for resource-constrained teams. This limits the framework's immediate applicability to organizations with established MLOps practices.

Fourth, metric selection and fairness definitions embed cultural and normative assumptions. Group fairness metrics (demographic parity, equalized odds) reflect Western individualist conceptions of fairness that may not align with collectivist cultural norms. The framework's adaptation pathways (Section~\ref{sec:adaptation-pathways}) address this partially, but the underlying metric choices remain culturally situated. Similarly, protected attribute definitions vary across jurisdictions, requiring careful mapping to local legal frameworks.

Fifth, environmental impact measurement—particularly for the Eco gate—carries significant uncertainty. Carbon footprint estimation depends on grid emission factors, data center PUE values, and hardware efficiency metrics that may be unavailable or imprecise. Water usage measurement is even more uncertain, as cooling water consumption varies with climate, facility design, and operational practices. These measurement challenges mean that Eco gate decisions may be based on noisy estimates, potentially leading to both false positives (blocking low-impact deployments) and false negatives (allowing high-impact deployments).

Finally, the framework's effectiveness depends on organizational buy-in and cultural change. Technical controls can be circumvented if organizational incentives reward speed over safety, or if governance processes lack enforcement authority. The framework addresses this through virtue-ethical emphasis on responsible practice and multi-layered accountability, but it cannot guarantee adoption or prevent organizational resistance to structural changes.

\section{Conclusion}

This paper has proposed a \textit{future-proof ethics-by-design framework} for NLP and IR that integrates consequentialist, deontological, and virtue-ethical reasoning into the AI pipeline from data collection to post-deployment monitoring. By bridging the principles-to-practice gap, the framework operationalizes philosophical commitments as measurable, enforceable lifecycle interventions. 

The framework's contribution lies in its operational control architecture: embedding moral reasoning, risk assessment, and accountability mechanisms as first-class elements of system architecture, rather than retrofitting ethics onto legacy pipelines. This AI-from-inception philosophy enhances resilience to paradigm shifts, from unimodal NLP to multimodal, agentic, and neuro-symbolic systems.

Future work should prioritize developing open-source tooling for continuous ethics auditing, refining metrics that remain robust across model classes, and establishing governance templates adaptable to varying legal and cultural contexts. For practitioners, the imperative is clear: ethics cannot be an afterthought; it must be an organizing principle from day one.

\backmatter



\section*{Declarations}
\begin{itemize}
\item Funding: This research received no funding.
\item Conflict of interest: The corresponding author states that there is no conflict of interest.
\item AI-assisted writing: Large language models were used for language polishing and structural editing during manuscript preparation. They were not used for conceptual development, metric design, threshold specification, or analysis. All framework design decisions, threshold choices, and analytical content were determined by the author.
\item Citation: Catapang, J.K. Building the ethical AI framework of the future: from philosophy to practice. AI Ethics 6, 150 (2026). https://doi.org/10.1007/s43681-026-01003-8
\end{itemize}

\begin{appendices}

\appendix
\renewcommand\thesection{\Alph{section}}
\section{Preregistered evaluation protocol}
\label{app:prereg}

\subsection*{A.1 Objectives}
\begin{enumerate}
  \item Estimate gate false-positive and false-negative rates by stage and metric.
  \item Quantify pre-release risk reduction and runtime detection performance relative to baseline.
  \item Characterize emissions–quality–latency trade-offs under multiple eco budget regimes.
\end{enumerate}

\subsection*{A.2 Endpoints}
\begin{enumerate}
  \item Primary endpoints:
  \[
    \mathrm{FP}_{\mathrm{gate}}=\Pr(\text{block} \mid \text{safe}), \qquad
    \mathrm{FN}_{\mathrm{gate}}=\Pr(\text{pass} \mid \text{risky})
  \]
  reported per stage and metric with 95\% bootstrap confidence intervals.
  \item Secondary endpoints:
  \begin{itemize}
    \item change in numeric risk metrics from baseline (same pipeline configuration without gates, or historical decisions if gates were not previously implemented) to with-gates,
    \item time-to-detection for drift and harms,
    \item emissions per query and training footprint versus quality and latency across budget regimes.
  \end{itemize}
  \item Sample size justification: target \(N_{\text{pre}}=150\) and \(N_{\text{run}}=40\) provide sufficient power to detect FP/FN rates differing from null hypothesis (e.g., FP=0.05, FN=0.10) with 80\% power at \(\alpha=0.05\) assuming moderate effect sizes, accounting for stratification and multiple comparisons.
\end{enumerate}

\subsection*{A.3 Scope and systems}
\begin{itemize}
  \item Fix system under test, model family, deployment modes, and regions before data collection.
  \item Record these choices in an artefact manifest; document any later deviations.
\end{itemize}

\subsection*{A.4 Data sources and sampling}
\begin{enumerate}
  \item Pre-release changes: target \(N_{\text{pre}}=150\) items, stratified by change type (data refresh, alignment run, guardrail edit, configuration change) in proportions reflecting typical pipeline operations or equal allocation if proportions unknown. Record item\_id, stage, timestamp, description, owner, source. Analyze when target reached or at fixed calendar date (see A.8).
  \item Runtime incidents: target \(N_{\text{run}}=40\) items, stratified by severity (high, medium, low) in proportions reflecting incident distribution or equal allocation if unknown. Use the same metadata fields. Analyze when target reached or at fixed calendar date.
  \item Test suites (fixed before gate runs):
  \begin{itemize}
    \item data stage fairness slices,
    \item train/align toxicity set,
    \item deploy domain QA for hallucination,
    \item monitor drift histograms.
  \end{itemize}
  \item Inclusion and exclusion: include items relevant to the stage with required metadata; exclude and list items with missing mandatory fields. For partial data (some metrics missing but others present), use complete-case analysis per metric; document missingness patterns.
\end{enumerate}

\subsection*{A.5 Metrics and thresholds}
\begin{itemize}
  \item Data: demographic parity difference and data quality.
  \item Train/align: toxicity mean and robustness.
  \item Deploy: hallucination rate and misuse block-rate.
  \item Monitor: drift (KL) and harm incident rate.
  \item Governance checks: documentation completeness, lawful processing, human oversight, transparency controls, incident SLAs.
  \item Eco checks: training and serving emissions budgets and, where applicable, water-use constraints.
  \item Register exact thresholds and directions (max, min, eq) in a time-stamped configuration file prior to analysis.
\end{itemize}

\subsection*{A.6 Gate rules}
\begin{enumerate}
  \item A gate blocks if any configured metric breaches its threshold or a governance check fails.
  \item Decisions are pass, block, or throttle.
  \item Log metric values, thresholds, decision, evidence pointers, timestamps, and signer identifiers.
\end{enumerate}

\subsection*{A.7 Labeling and adjudication}
\begin{enumerate}
  \item Two independent reviewers label each item safe or risky with rationale and confidence.
  \item Compute Cohen's kappa prior to adjudication. If \(\kappa < 0.6\), conduct calibration discussion and re-label; if still below threshold after calibration, exclude item from primary analysis and report separately. If \(\kappa \geq 0.6\), apply majority vote; resolve ties by consulting a third independent reviewer (blinded to gate outcomes and prior labels) rather than defaulting to risky, to avoid conservative bias.
  \item Keep reviewers blinded to gate outcomes.
\end{enumerate}

\subsection*{A.8 Analysis plan}
\begin{enumerate}
  \item Estimation: compute false-positive and false-negative rates per stage and metric with 2000 bootstrap resamples and a fixed random seed.
  \item Baseline comparison: where paired baseline decisions exist (same items evaluated with and without gates, or matched historical controls), use McNemar's test for error profile differences.
  \item Numeric effects: summarize changes in risk metrics as means with bootstrap intervals.
  \item Multiple comparisons: report per-metric intervals; provide a Benjamini–Hochberg adjusted summary in appendix tables.
  \item Sensitivity: vary thresholds within pre-specified bands; for eco metrics vary grid intensity and cooling-water assumptions within declared ranges.
  \item Stopping and missing data: analyze when targets (\(N_{\text{pre}}=150\), \(N_{\text{run}}=40\)) are reached or at a fixed calendar date (specified before data collection begins). For missing data: use complete-case analysis per metric (exclude items missing that specific metric but include for other metrics); document missingness patterns and report counts of excluded items per metric. Omit unlabeled items (those failing inter-rater reliability threshold) from primary analysis and report separately.
\end{enumerate}

\subsection*{A.9 Robustness and threats to validity}
\begin{itemize}
  \item Construct validity: spot-check metric computations against raw artefacts.
  \item External validity: results apply to sampled domains and languages; state limits.
  \item Governance verification: verify checks against artefact hashes.
  \item Known limitations: enumerate and discuss in the final report.
\end{itemize}

\subsection*{A.10 Reporting and artefact release}
\begin{itemize}
  \item Release configuration files, evaluation scripts, test suites or generators, gate logs or redacted summaries, and analysis notebooks required to regenerate tables.
  \item Include checksums and a license; exclude user-identifying content.
\end{itemize}

\subsection*{A.11 Deviations}
\begin{itemize}
  \item Document any deviation from the protocol, including threshold changes after outcome inspection, with justification and time stamp.
\end{itemize}

\subsection*{A.12 Results policy}
\begin{itemize}
  \item No results are reported in this manuscript version; executing the protocol will populate the tables referenced in the main text.
\end{itemize}




\end{appendices}


\bibliography{sn-bibliography}

\end{document}